# Influence of Protein Electrostatic Field on Hydrogen Bonding

S. Mayburov[1*], C. Nicolini[2,], V. Sivozhelezov[2,3]

[1]*Lebedev Inst. of Physics, Leninski pr. 53, Moscow, Russia, 117924 E-mail: mayburov@sci.lebedev.ru*

[2]*Nanoworld Inst. and Biophysics Chair, University of Genoa Corso Europa 30, 16132 Genoa, Italia E-mail: manuscript@ibf.unige.it \*, vsivo@nwi.unige.it*

[3]*Institute of Cell Biophysics, Russian Academy of Sciences, 3 Institutskaya, Pushchino, Moscow Region 142290 E-mail: vsivo@icb.psn.ru*

\*Corresponding Author

## Abstract

The quantum-mechanical mechanisms by which the enzymes catalyze the hydrogen transfer in biochemical reactions are considered. Up to date it was established both experimentally and theoretically that in many cases the proton tunnelling through the intermolecular potential barrier is essential. We argue that in this case the enzyme excitation and internal motion facilitate proton transfer between reactants by squeezing the potential barrier which otherwise is practically impenetrable. In the similar fashion, the enzymes can facilitate the formation of hydrogen (H) bonds between the molecules. By means of barrier squeezing, the enzymes not only facilitate such reactions but also can control their rate and their final outcome, depending of enzyme excitation. In particular, such effects can play the major role in DNA polymerization reactions where preliminary DNTP selection is quite important.

## 1. Introduction

The microscopic mechanisms by which the enzymes catalyze biochemical reactions to achieve enormous rates are not well studied up to date. However, the recent investigations indicate that at least for some relatively simple processes the principal role play the transitions through the classically forbidden states, in particular, the proton tunnelling for hydrogen transfer reactions (Hammes-Schiffer, 2006). We now briefly recall the aspects of the proton tunnelling which are important for biochemical reactions catalyzed by enzymes. The quantum tunnelling, i.e. the passage of quantum particles under the potential barrier is the direct consequence of quantum-mechanical (QM) uncertainty principle. To illustrate it, consider the free particle of mass $m$ with kinetic energy $E$ directed along the coordinate axe $X$ to the potential barrier $V_B(x)$. If $E > U$, where $U=max(V_B)$, then both the classical and quantum particle would pass over it effectively, yet in quantum case even if $E < U$, some fraction of the flow of particles $T_0$ will penetrate through the barrier and can be observed on its opposite side; $T_0$ is called the tunnelling (transmission) coefficient. Solving 1-dimensional Schrodinger equation and neglecting small terms, it follows:

$$T_0 \cong e^{-\frac{2}{\eta}S} = e^{-\frac{2}{\eta}\int\sqrt{2m[V_B(x)-E]}dx} \qquad (1)$$

here $S$ is WKB action, $\eta$ is Planck constant (Landau, 1974). For the rectangular barrier of height $U$ and width $l$, it gives:

$$T_0 = e^{-\frac{2l}{\eta}\sqrt{2m(U-E)}} \qquad (2)$$

In particular, the importance of proton tunneling for DNA stability and transfer of genetic code was discovered in the sixties (Lowdin,1963). It was shown that the proton tunneling between the nearby p-donors or p-acceptors is the potential source of DNA mutations (Kryachko, 2003). Such DNA mutations can occur via the proton tunneling inside the same nucleotide or between two nucleotides, they called tautomerisation (Li,2001; Dabrowska, 2004).

The hydrogen (H) bonds are formed between the p-donor and p-acceptor atoms or molecular groups, designated below both as pDA. Usually, p-acceptor atoms possess lone pairs of electrons on their external shell; this is typical for electronegative atoms like N,O,F, etc. The p-donors are also mainly electronegative atoms which form stable chemical bonds with H. If such molecular group and p-acceptor approach closely to each other, the proton can penetrate through the potential barrier $V_B$ between them to the location of p-acceptor atom. When the stable H-bond is formed, the proton state is the superposition of its states in potential wells of donor and acceptor. In other words, the proton oscillates between the donor and acceptor regions. As the result of proton delocalisation, the donor and acceptor groups attract one another electrostatically with the effective charges of the order $eP_{1,2}$, where $P_{1,2}$ are the proton localization probabilities in donor and acceptor regions. In addition, the fermionic exchange interactions produce the minor repulsion effect, other terms are relatively small. Our model exploits the potential approximation, in which H-bonding is described by the proton stationary state in the double well potential (Joesten, 1974). H-bonds are quite sensitive to the electrostatic environment and presence of other H-bonds. In particular, in DNA double helixes, each H-bond cannot be properly described without account of other nearby H-bonds (Lowdin, 1963).

## 2. Methods and Results

The kinetics of H-bond formation between the proteins is far from exhaustively studied, but some experimental results suggest that the proton tunneling between p-donors and p-acceptors in vivo can be one of its main mechanisms. For the inorganic molecules, in particular, $H_2O$ dimers and chains, the proton kinetics is successfully described as the semiclassical proton transfer between p-donors and p-acceptors (Davidov, 1991). The computer simulations and experiments show that some DNA mutations also start from the proton tunneling between the neighbour nucleotides and result in H-bonds formation in tautomers (Lowdin, 1963, Kryachko, 2003). Basing on these facts, the approach adopted herein implies that H-bonds formation also starts from the proton tunnelling between p-donor and p-acceptor. Indeed, to start H-bond formation in a ground state, the proton should tunnel first through the molecular potential barrier to p-acceptor, but if this barrier $V_B$ is too large, i.e. $T_0 \to 0$, it becomes less and less probable. Plainly, if H-bond formation occurs via the proton tunnelling, then the rate $R_H$ of H-bond formation should grow with $T_0$ of (1). In addition, if the barrier enlarges, the proton will be located near p-acceptor with essentially smaller probability $P_2$, so the electrostatic attraction between p-donor and p-acceptor will be too weak to form the stable H-bond. In addition to these calculations, our main results will be derived by the alternative method, which analyses the stationary H-bonding states and does not exploit the hypothesis of initial tunnelling directly.

The majority of biochemical reactions between the reactant molecules (substrates) $Re_i$ practically can only proceed in the presence of enzymes $En$, the typical reaction scheme in standard theory of catalysis i.e. Transition State theory is :

$Re_1 + Re_2 + En \to X_R \to En + Pr_1 + Pr_2 + \ldots$  (3)

where $X_R$ is the transition state, $Pr_i$ are the reaction products. It was found recently that some enzymes reduce the width of proton potential barrier between reactants; it results in much higher rate for the hydrogen transfer reactions than standard theory predicts (Masgrau, 2006). The physical mechanism of this effect is still disputed, but there are the strong indications that the p-barrier width depends on the internal state of enzyme, in particular, of its internal protein motion (Hammes-Schiffer, 2006). For organic molecules, the reactions resulting in H-bonds formation, are similar to the proton transfer reactions in many aspects (Joesten, 1974). Hence the regulation of proton tunneling considered herein can be important for them as well, in particular, for H-bonding between nucleotides in DNA.

Currently, the influence of enzymes on the proton tunneling is described by the phenomenological theories, in which the proton transfer rate affected mainly by the thermal fluctuations of enzyme shape (Bruno, 1992, Hammes-Schiffer, 2006). It supposed that the dependence of the potential barrier width $l$ on the enzyme state can be described by the dynamics of harmonic oscillator, so that $l$ is controlled by the oscillator $O_E$ with the potential $U_O(l) = \frac{k}{2}(l - l_{eq})^2$ where $l_{eq}$ is its value at the equilibrium. In this case, the thermal fluctuations of $O_E$ energy would result in the variations of $l$ value, and accordingly, the average tunneling coefficient of (2) would rise significantly.

Another regime, probably more interesting for the reactions of DNA polimerization, is the one for which main role in $l$ variations play the collective excitation of enzyme. We shall consider here such $l$ variations of two kinds: oscillating and quasistatic ones. In the oscillating mode, the ordered excited state $G$ can be described as a shock wave (phonon) with constant energy of oscillator $E_O$ which results in the constant amplitude $\Delta$ of $l$ oscillations during the excitation period $\tau$. If the thermal and other $l$ fluctuations are much less than $\Delta$, then $l(t) = l_{eq} + \Delta \cos \omega t$, i.e. the potential barrier for proton $V_B$ will be time dependent. As the result fro the rectangular barrier the tunnelling coefficient averaged over time is equal to:

$$\overline{T} = T_0 \frac{1}{\tau} \int_0^\tau dt\, e^{-\frac{2\Delta \cos \omega t}{\hbar}\sqrt{2m(U-E)}} \qquad (4)$$

where $\omega$ is $O_E$ frequency, $T_0$ value is given by (2) for $l=l_{eq}$. $U \approx .3 \div .5$ eV, $l_{eq}=.5$ Å are the typical parameters of p-barrier for H-bonds formation. The kinetic energy of proton $E$ is much less than $U$, hence $E=0$ can be taken in the calculations, and then $\overline{T}$ can be easily calculated analytically. However, it is not clear whether the effective formation of H-bonds is possible in time-dependent proton potential $V_B(t)$ which oscillates quite fast. Besides, in such model the average tunnelling coefficient $\overline{T}$ can only increase over its value for the ground state, yet we are interested also in the case when it decreases. Hence the quasistatic variant of excitation, when during the period $\tau$ the potential barrier $V_B$ is constant, yet its form and/or magnitude can change promptly at the instant $t=0$, seems to be more promising. For example, if only the width of rectangular barrier changes, then the excitation $G_i$ results in the reduced (or enlarged) barrier width $l(t)=l_{eq} \pm \Delta$ for time $0<t<\tau$. For the same values of parameters as in oscillating case, for $\Delta=.1$ Å one obtains that $T_0$ of (2) changes by a factor 10, in comparison with the initial $T_0$ value which is about $10^{-6}$. As will be argued below, for the electrostatic mechanism of barriers regulation, the quasistatic variant seems to be more realistic, so we will further considered here only this variant, while the oscillating one will be studied in our forthcoming paper. It will be also shown below that for quasistatic mode, the influence of thermal fluctuations can be neglected.

Now let's discuss the possible microscopic mechanism which permits the enzyme to regulate the potential barriers between reactants. Until now this issue has not been fully clarified, but there are the indications that the electrostatic effects can play the important role in it (Benkovic, 2003). Most of enzymes have complex shapes and are comparatively large at molecular scale, so the electric polarisation (Dolan et al., 2004) and other electrostatic phenomena such as focusing (Sheinerman, 2000) inside them are possible. Note that H-bonds are also quite sensitive to their electrostatic environment. The experiments with the inorganic molecules show that the binding affinity of H-bond would increase 4-6 times, if the analogous complex is anionic. Such H-bonds called ionic, the example being $(HF)_6$ versus $[F…H…F]^-$ (Joesten, 1974).

In the same vein the recent experiments and computer simulations for DNA nucleotides and their molecular analogs show the strong influence of electric charges configurations on proton transfer (Dabkowska,2004). The most important of them is the observation of barrier-free proton transfer (BFPT) in nucleotide dimers: if one nucleotide is anionic, then the rate of proton transition from other nucleotide or its analogue shows the practical absence of any potential barrier, whereas such barrier is large for the same but neutral dimer (Gutowski,2002; Li 2001). The examples are anionic Uracyl, Glycine dimer; the other one most interesting for us is $C^-$,G dimer. In that case BFPT obtained for proton bound initially to N atom of G and attached finally to N atom of C anion (Li,2001). The calculated barrier height is about .05 eV and in this case it's about the value of proton kinetic energy at given conditions, whereas for neutral G,C dimer this p-barrier is about .6 eV and width .5 Å (Zoete,2004). Basing on these results, we propose the electrostatic mechanism of proton barrier regulation for the control of H-bond formation rate between DNA-pol and dNTP. The investigation of DNA-pol conformations shows that during the induced-fit the electrostatic configuration of DNA-pol changes significantly, and such changes influence directly on some features of DNA replication (Li, 1998, Kunkel, 2004). In particular, the electric charges and dipoles, which appear near DNA-pol surface, can influence the rate of H-bonds formation with dNTP.

For illustration, consider the situation when only one H-bond should be formed between the binding site and the free floating molecule which, for example, can be p-donor $A_P$. In our model H-bond formation supposedly begins from the transition of proton in the ground state from $A_P$ potential well to the potential well of acceptor $A_A$, which are divided by the barrier $V_B$ with the height about .6 eV (Zoete, 2004). Assume that the negative charge $q$ is located under DNA-pol surface, its electrostatic potential is $V_E(\vec{r}) = -\frac{q}{r}$, where the coordinate $\vec{r} = 0$ in the charge centre and axe $X$ is directed orthogonally to DNA-pol surface. The potential $V_E$ has the sign opposite to that of $V_B$, the total proton potential is $V_T = V_B + V_E$, and for relatively large $q$ and small $r$ the potential $V_E$ can effectively reduce the barrier for proton transfer between the donor and acceptor. This modified potential will enhance the rate of proton transfer to DNA-pol, and could stimulate H-bond formation, if p-acceptor is located on DNA-pol surface nearby to $q$. Accordingly, if $q$ is positive, its field will hinder H-bond formation in such system, i.e. it corresponds to the effective enlargement of proton potential barrier. Such process is the key element of H-bonding regulation in our model, its rate will be calculated here for typical nucleotides parameters. Let's suppose that DNA-pol surface lays at the distance $d$ from $q$ centre and is locally flat; p-acceptor $A_A$ is located on DNA-pol surface at $y=z=0$. The rectangular p-barrier $V_B$ of height $U$ and width $l$ starts at the distance $x_B$ from $q$ centre; the proton with kinetic energy $E$ moves from the direction $x \to \infty$ (fig.1). For simplicity, the calculations are first done in 1-dimensional approximation, i.e. we take $V_E(\vec{r}) = \frac{q}{x}$, which is reasonable for the region around the point $y=z=0$ and at large distance from $q$; the corrections introduced by 3-dimensional case will be accounted below. For potential of such complex form the WKB action of (1) is equal to:

$$S = \int_{d}^{x_{max}} \sqrt{2m[V_T(x) - E]} dx = \int_{d}^{x_{max}} \sqrt{2m[\frac{eq}{x} + V_B(x) - E]} dx, \quad (5)$$

where the value of $x_{max}$ defined below, but it larger than $x_B+l$. The integral can be easily calculated analytically, but the exact ansatz, which is rather tedious, is omitted here because in practical cases it allows the simple approximation. It gives value $T_0$ of (1), which in semiclassical approximation is the probability for proton to reach p-acceptor located at the distance $d$ from $q$ centre.

To estimate the possible level of H-bonding suppression one should compare $T_0$ of (1) with and without the charge presence for typical dNTP parameters. We take that $q=|e|$, i.e. this is the single ion, and choose arbitrarily $d=2$ Å. For the proton barrier between nucleotides the reasonable approximation is $U$ value .5 eV, $l=.5$ Å (Zoete, 2004); the realistic estimate for the width of p-acceptor potential well is about 1 Å, hence $x_B=3$ Å. The tunnelling proton is not free initially, but confined in p-donor well and moves together with the molecular centre of mass in the direction of DNA-pol surface. The average kinetic energy $E$ of transferred proton is quite small, $E$ distribution is described by Boltzman spectra for room temperature (Hammes-Schiffer, 2006). Hence $E=0$ is assumed in our calculations, and the role of initial binding manifests mainly in the value of $x_{max}$, which corresponds to the average position, from which the proton starts its transition to p-acceptor, when dNTP molecule approaches DNA-pol surface. The reasonable estimate of its distance to the barrier edge is about 1.5 Å, so that $x_{max}=5$ Å (fig.1). For such choice of parameters, substituting $S$ of (5) into $T_0$ of (1) shows that the appearance of charge $q$ reduces the $T_0$ value by a factor $10^{40\div50}$, and that obviously makes p-tunneling and H-bonds formation practically impossible. The thermal fluctuations of proton energy $E$ do not change this results significantly. Note that in this range of parameters $S$ of (7) can be decomposed with good accuracy as:

$$S \cong \int_{d}^{x_{max}} \sqrt{2m(\frac{eq}{x} - E)} dx + \frac{1}{2}(U - E) \int_{x_B}^{x_B+l} dx \sqrt{\frac{2mx}{eq}} \cong 2\sqrt{2meq}(\sqrt{x_{max}} - \sqrt{d}) \quad (6)$$

where the first integral is much larger than the second one. Overall, $S$ value in such ranges of parameters shows the weak dependence on $d$ and $x_B$ values and slightly stronger on $x_{max}$, so the obtained $S$ estimates are relatively stable and the slight variations of parameters would not change our main result. Note that if the charge $q$ is removed further from DNA-pol surface, then as follows from (5), $S$ will be reduced proportionally to $d^{-\frac{1}{2}}$, i.e. H-bond formation can be regulated effectively by such $q$ shift.

Now let's consider the corrections induced by the exact form of $V_E$ in 3 dimensions. To calculate them exactly, solution of 3-dimensional Schroedinger equation is required, which is quite difficult in the considered configuration, so only estimates will be presented here. We suppose that the transverse radius of the surface through which the proton spreads during tunnelling $r_T=1$ Å. To get the upper limit of $T_0$ of (1) one can calculate $S$ on the cylindrical surface of tunnelling region, where $S$ is minimal, for the integration path along this surface it expressed as:

$$S = \int_d^{x_{max}} \sqrt{2m[\frac{eq}{\sqrt{x^2+r_T^2}} + V_B(x) - E]} dx \qquad (7)$$

This integral presented analytically as the sum of elliptic integrals, the calculations show that its value differs from 1-dimensional case by less than 10%, and so cannot change our results principally. Plainly, the same suppression effect will be observed for p-donor on DNA-pol surface and the negative charge located near it.

The similar effect of H-bonds suppression follows from the considerations of the stationary H-bonding system in Coulomb field. As was mentioned above, in the potential approximation, the stable H-bond is described by the proton stationary state with energy $E_B$, extended inside two potential wells. Such double well is asymmetric, for nucleotides the depth of p-donor well is about .6 eV, and is about .4 eV for p-acceptor well, $E_B$ value is about .3 eV (Joesten, 1974). Under this conditions the probability $P_2$ of proton localization in acceptor well is about $10^{-1}$, which results in a sizeable electrostatic attraction between p-donor and p-acceptor. In the considered example, however, the external Coulomb field rises the edge of acceptor well about 1.5 eV higher than the edge of p-donor well, i.e. its bottom will be essentially higher than $E_B$ level. Then, for the considered external field any stationary state of proton in one of wells will be exponentially suppressed inside another. To demonstrate it, let's consider the stationary state of proton with energy level $E_B$ in the p-donor well. In such external field, the probability to find the proton in the acceptor well is estimated as $P_2 \approx e^{-\frac{2R}{\eta}}$ (Landau, 1972), where $R$ is equal to:

$$R = \int_{\frac{1}{2}(d+x_B)}^{x_b+l} \sqrt{2m[V_T(x) - E_B]} dx \qquad (8)$$

For our parameters, it gives value of $P_2$ about $10^{-15}$, which excludes any sizeable electrostatic attraction between p-donor and p-acceptor. Hence, there are no stationary proton states in such electric field, i.e. the stable H-bond between p-donor and p-acceptor cannot be formed at all. Summing up, the considerations of stationary proton states in potential approximation confirms the conclusions which follows from the kinetics of proton tunnelling.

The physical meaning of obtained results is quite obvious: the proton with low kinetic energy $E$ cannot approach closely to positive ion, which effective coulomb potential on DNA-pol surface is about 3 eV in this set-up. But even if $E$ will be high enough to reach the p-acceptor location, the external electric field will not permit to form the stable H-bond, because the location probability $P_2$ will be too small to result in the essential electrostatic interaction of p-donor and acceptor. Such significant influence of elementary charge on H-bonds formation is easy to understand if we recall that the relative strength of H-bonds and electrostatic ionic bonds is about $10^{-1}$ (Joesten, 1974). The considered suppression of H-bonds formation can be proved more straightforwardly by means of computer simulations, but even such simple estimates are quite illustrative.

The analogous calculations can be performed for the electrostatic field of dipole located nearby to DNA-pol surface. Suppose that the dipole centre is at $\vec{r} = 0$ and consists of positive and negative ions at the distance 2 Å, aligned along $X$ axe with resulting dipole moment $\vec{P}_Q$. Substituting the dipole potential $V_D(\vec{r}) = \dfrac{\vec{P}_Q \vec{r}}{r^3}$ into $V_T$, one obtains in the 1-dimensional approximation:

$$S = \int_d^{x_{\max}} \sqrt{2m[V_T(x) - E]}\,dx = \int_d^{x_{\max}} \sqrt{2m[\dfrac{e|\vec{P}_Q|}{x^2} + V_B(x) - E]}\,dx, \qquad (9)$$

which is easily expressed analytically. It follows that in this case the principal effect of tunnelling suppression will be as large as for the single charge of (5).

In these calculations we neglected the possible rearrangement of nearby atomic shells and the polarization of surrounding media induced by the charge $q$. However, we do not expect that these effects can change significantly the obtained results; due to quite small distance between DNA-pol and dNTP, the polar molecules of $H_2O$ should be mainly outside of tunnelling region. In addition, DNA-pol surface possess the significant hydrophobic properties (Kunkel, 2000). The obtained results are applicable for the case of single H-bond formation, yet dNTP posses 2 or 3 H-bond vacancies which can be bound with the particular binding site simultaneously. Hence their confinement by the binding site of dNTP-pol would be performed via the tunnelling of 2-3 protons via the different barriers. Such double proton transfer was already observed experimentally for some organic molecules (Limbach et al., 2004); the computer simulations confirm its existence for G,C nucleotide dimer (Zoete, 2004). Hence to prevent the binding of dNTP with two or three H-bonds vacancies via such mechanism, the binding site should include the array of positive and negative charges located near DNA-pol surface, so that in excited state they are removed and all p-barriers are reduced simultaneously to the form and size, optimal for H-bonds formation. The possible variants of such pDA configurations that form specific binding sites will be considered elsewhere.

It cannot be excluded also that the analogous electrostatic effects, beside the suppression of H-bonding, can stimulate H-bonds formation between dNTP and DNA-pol also, as was proposed above. However, the calculations of such effects demand the complicated computer simulations and are not considered here. Our estimates of such proton transfer enhancement were performed analogously to calculations of (6) – (8), they show that, in this case, the effect will be probably less pronounced, than for p-transfer suppression. Overall, it seems that the electrostatic structures permit many different opportunities for the control and regulation of dNTP binding by DNA-pol. Note that the analogous mechanism of electrostatic blocking regulates the transfer of the metal ions through the cell membranes (Davidov, 1991). This considerations of electrostatic effects does not exclude the alternative mechanisms of potential barrier reduction which act simultaneously with the electrostatic one. Further, we only discuss this electrostatic mechanism, because it seems most simple and appropriate for the considered process. In addition, the electrostatic mechanism of tunnelling regulation is quite fast and practically reversible.

## 3. Discussion

Recent experiments provide evidence that the variation of p-tunneling in some biochemical reactions induced electrostatically at the microscopic level (Benkovic, 2003; Masgrau, 2006 ). In general, both the dipole moments on enzyme surface and the thermal fluctuations of charge density inside its volume can produce the stochastic electric field. Due to the large size of enzymes its average magnitude can achieve a significant strength. If such field has the suitable orientation, it will suppress or enlarge the initial proton barriers between molecules (Hammes-Schiffer,2006). In our approach, the necessary effect is caused by the regular field of ions, while the thermal fluctuations of such field are neglected, because the estimates show their relative smallness.
In the model framework it was shown , in particular, that the electric field of positive ion located near p-acceptor on DNA-pol surface can effectively block the formation of H-bond with dNTP p-donor. For the typical nucleotide parameters the tunnelling of proton $T_0$ is suppressed by a factor of $10^{-40}$, the effective attraction between donor and acceptor is reduced by a factor $10^{-15}$, it practically excludes H-bonds formation between dNTP and DNA-pol. Here only the effects of very simple electrostatic configurations were estimated, the computer simulations should help to define those and other quantitative aspects of our model more precisely.

Basing on these calculations, we shall discuss now the possible role of such electrostatic regulation in the reaction of DNA polymerization (replication). To prepare the exact copy of initial DNA, dNTPs should be identified correctly before to be implanted as the partner of given base template in a DNA strand. This operation is characterized for some DNA-pols by unexpectedly high level of fidelity $F$, i.e. the percentage of wrong nucleobases in constructed DNA structure. The chemical kinetics calculations predict $F$ of the order $10^{-1} \div 10^{-2}$ , whereas, the experiments with replicative DNA-pols show that after the implantation of nucleotide $F$ value can achieve the level of $10^{-5} \div 10^{-6}$; the additional $F$ improvement by a factor 10 is achieved by a secondary check by exonuclease (Beard, 2002; Kunkel, 2004).

The huge diversity of DNA-pols was found in the last years, they constitute several large families designated as A,B,Y, etc. (Lodish, 2000). The members of different families can have the fidelity differing by many orders, while some types of DNA-pols perform also the repair of DNA lesions (Beard, 2002; Vaisman, 2005). For the majority of DNA-pols its geometric form is analogous to the human right hand and can be divided into the palm, thumb and fingers domains, the fingers supposedly interact first with incoming dNTP. The experiments evidence that during DNA replication DNA-pol passes through the complex sequence of conformational changes, which include the mechanistic movement of some domains or their parts (Beard, 2002). Normally, the palm is open, but when incoming dNTP approaches the fingers domain of DNA-pol, some parts of fingers and thumb domains rotate and DNA-pol 'fist' closes, driving two captured nucleobases into the close contact – so called the 'induced-fit' transformation (Sawaya, 1997). As the result, the final form of DNA-pol provides the optimal conditions for base pair polymerisation, since in this conformation DNA-pol domains constitute the closed pocket, in which base pair is confined tightly. Some experiments evidence that the 'induced-fit' occurs practically only if the incoming dNTP is complementary to DNA base template, suggesting some preliminary selection of base pairs (Kunkel, 2004; Purohit, 2003). In case of incorrect base pairing, DNA-pol returns into its initial state in a short time, i.e. the induced-fit is a substrate dependent transformation (Post, 1995, Purohit, 2003). Despite the large amount of experimental information obtained in the last years, the detailed mechanism of the induced-fit initiation is not clear till now (Kunkel, 2004).

The experiments suggest that the observed fidelity $F$ of dNTP selection is defined mainly by the final selection of base pairs performed when DNA-pol is in closed conformation (Kunkel, 2004). For the majority of DNA-pol types this final dNTP selection performed mainly according to the size and shape of base pair – the steric effect, yet for some DNA-pols the structure of dNTP hydrogen (H) bonds serves the main identification mark (Kool, 2001; Wolfe, 2005). Here we proposed the model, in which dNTP preliminary selection also is owed to the different structure of dNTP H-bonds. The nucleobases posses the different geometric configurations of their p-donors and p-acceptors, due to this distinctions, the specific molecular groups on DNA-pol surface supposedly can bind strongly only dNTPs of one particular kind. It can be supposed that such groups are forming specific binding sites (sometimes termed as the hot spots) and activated by the specific excitations of DNA-pol, which depend on the type of DNA base template. It's shown that for hot spots of particular molecular structure only dNTP complementary to the base template can be captured by DNA-pol and transferred to its catalytic cite. In the approach developed herein, the mechanism of hot spots activation has the electrostatic origin. Namely, the ions, located nearby to p-donors or p-acceptors of hot spots, block the proton transfer by their electrostatic potential. DNA-pol excitation remove this charges from the hot spot far enough, so that the potential barriers are reduced and protons can pass to or from dNTP with high efficiency and form H-bonds. It seems possible that such transfer of protons occurs in vivo mainly via the proton tunnelling through the molecular potential barriers.

## Figure Captions

Fig.1.   The total proton barrier potential  $V_T$  produced by positive ion  $q^+$  and molecular proton barrier  $V_B$  near DNA-pol surface

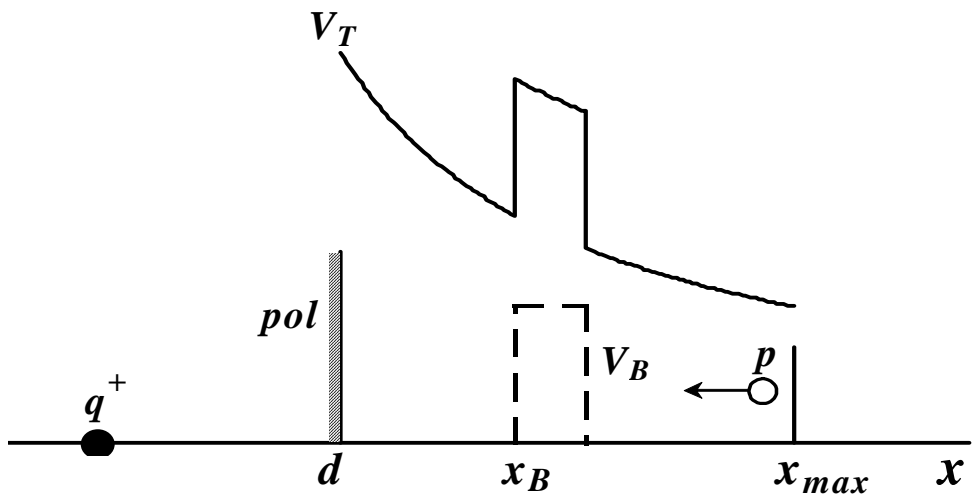

Fig.1